\documentclass[a4paper]{jpconf}
\usepackage{graphicx}

\newcommand{\ba}{\begin{eqnarray}}
\newcommand{\ea}{\end{eqnarray}}
\newcommand{\beqs}{\begin{eqnarray}}
\newcommand{\eeqs}{\end{eqnarray}}

\begin{document}
\title{Momentum transfer dependence of the hadron GPDs
       and Compton form factors} 

\author{O.V. Selyugin}

\address{ BLTPh, JINR,  Dubna, Russia}

\ead{selugin@theor.jinr.ru}

\begin{abstract}

    The generalized parton distributions (GPDs) of the meson and nucleon at small and large values
    of the momentum transfer were determined
   on the basis of the comparative analysis of different sets of  experimental data
    of electromagnetic form factors of the proton and neutron,
      using the different sets of the parton distribution functions (PDFs).
    As a result, different form factors of the nucleons and meson were calculated.
    The t-dependence of these form factors
    was checked up in the description of the nucleon-nucleon and meson-nucleon elastic  scattering
    and the differential cross sections of the real Compton scattering.

\end{abstract}

\section{Introduction}

 The recent results from the LHC  gave  plenty of new information
  about the elastic and dip-inelastic  processes which leads to   new questions in the study of the structure of hadrons.
   At first, it is needed to note that the new data of the TOTEM and ATLAS Collaborations on the elastic proton-proton scattering
    indeed show that none of the old model predictions
    give the true descriptions of the elastic cross sections at the LHC  \cite{Rev-LHC}.
              The different reactions   can be related with the different form factors.
     The strong hadron-hadron scattering can be proportional to the matter distribution of the hadrons.
       The Compton scattering is described by the Compton form factors.
     Now we have  more general parton distributions which depend on  different variables -
     $GTMDs(x,\vec{k},\xi,\vec{\Delta)}$ - Generalized transverse momentum dependent parton distributions.
     \cite{Meis-08,Lorce-13,Burk-15}.    They parameterize the unintegrated off-diagonal quark-quark correlator,
     depending on the three-momentum $\vec{k}$ of the quark and on the four-momentum $\Delta$, 
     which is transferred by the probe to the hadron. Taking $\Delta=0$,  we can obtain the
     $TMD(x,\vec{k})$ - transverse momentum-dependent parton distribution. In other way,
     after integration over $\vec{k}$ we obtain $GPDs(x,\xi,\Delta)$ - Generalized parton distributions.
     The remarkable property of the $GPDs$ is that the integration of the different momentum of
     GPDs over $x$ gives us the different hadron form factors  \cite{Mil94,Ji97,R97}.

    The $x$ dependence of $GPDs$ in most part is determined by the standard
    PDFs, which are obtained by the different Collaborations from the analysis of
    the dip-inelastic processes.
 Many different forms
  of the $t$-dependence of GPDs were proposed.
   In the quark diquark model \cite{Liuti1} the form of  GPDs
   consists of three parts - PDFs, function distribution and Regge-like.
 In other works (see e.g. \cite{Kroll04}),
  the description of the $t$-dependence of  GPDs  was developed
  in a  more complicated picture using the polynomial forms with respect to $x$.

 \section{Momentum transfer dependence of GPDs }

Let us modify the original Gaussian ansatz
 and choose  the $t$-dependence of  GPDs  in a simple form
$ {\cal{H}}^{q} (x,t) \  = q(x) \   exp [  a_{+}  \
   f(x) \ t ],                                     $
  with $f(x)= (1-x)^{2}/x^{\beta}$
   \cite{ST-PRDGPD}.
The isotopic invariance can be used to relate the proton and neutron GPDs.

   The complex analysis of the corresponding description of the electromagnetic form factors of the proton and neutron
    by the different  PDF sets  (24 cases) was carried out in \cite{GPD-PRD14}. These
   PDFs include the  leading order (LO), next leading order (NLO) and next-next leading order (NNLO)
   determination of the parton distribution functions. They used the different forms of the $x$ dependence of  PDFs. 
    We slightly complicated the form of GPDs  in comparison with the equation used in     \cite{ST-PRDGPD},  
   but it is the simplest one as compared to other works (for example \cite{DK-13}).
\ba
{\cal{H}}^{u} (x,t) \  = q(x)^{u}  \   e^{2 a_{H}   f(x)_{u}  \ t };  \ \ \ 
{\cal{H_d}}^{d} (x,t) \  = q(x)^{d}  \   e^{2 a_{H} f_{d}(x)  \ t };  \\
{\cal{E}}^{u} (x,t) \  = q(x)^{u} (1-x)^{\gamma_{u}} \   e^{2 a_{E}  \  f(x)_{u}  \ t }; \ \ \  
{\cal{E_d}}^{d} (x,t) \  = q(x)^{d}  (1-x)^{\gamma_{d}} \   e^{2 a_{E} f_{d}(x) \ t },
\label{t-GPDs-E}
\ea
 where
 $ f_{u}(x) =  \frac{(1-x)^{2+\epsilon_{u}}}{(x_{0}+x)^{m}}$ and $f_{d}(x) = (1+\epsilon_{0}) (\frac{(1-x)^{1+\epsilon_{d}}}{(x_{0}+x)^{m}} )$.

 The hadron form factors will be obtained  by integration  over $x$ in the whole range $0 - 1$.
 Hence, the obtained  form  factors will be dependent on the $x$-dependence of the forms of PDF at the ends of the integration region.
 The different Collaborations determined the  PDF sets  from the inelastic processes only in  some region of $x$, which is only
 approximated to $x=0$ and $x=1$.
   Some  PDFs  have the polynomial form of $x$ with
     different power.  Some other have the exponential dependence of $x$.
  As a result, the behavior of  PDFs, when $x \rightarrow 0$ or $x \rightarrow 1$,  can  impact  the
    form of the calculated form factors.

    On the basis of our GPDs with
    ABM12 \cite{ABM12} PDFs    we calculated the hadron form factors
     by the numerical integration
   and then
    by fitting these integral results by the standard dipole form with some additional parameters
$   F_{1}(t)  = (4m_p - \mu t)/(4m_p -  t ) \  \tilde{G}_{d}(t) $
  with $ \tilde{G}_{d}(t) = 1/(1 + q/a_{1}+q^{2}/a_{2}^2 +  q^3/a_{3}^3)^2 $ which is slightly  different from
  the standard dipole form on two additional terms with small sizes of coefficients.
  The matter form factor 
\ba
 A(t)=  \int^{1}_{0} x \ dx
 [ q_{u}(x)e^{2 \alpha_{H} f(x)_{u} / t  } 
  + q_{d}(x)e^{ 2 \alpha_{H} f_{d}(x)  / t}  ] 
\ea
 is fitted   by the simple dipole form  $  A(t)  =  \Lambda^4/(\Lambda^2 -t)^2 $.
        These form factors will be used in our model of the proton-proton and proton-antiproton elastic scattering.

\section{Magnetic transition form factor  $G^{*}_{M (\gamma^{*} N \Delta ) } $    }

  To check the momentum dependence of the spin-dependent part of GPDs $ E_{u,d}(x,\xi=0,t) $, 
   we can calculate the magnetic transition
  form factor which is determined by the difference of $ E_{u}(x,\xi=0,t) $ and $ E_{d}(x,\xi=0,t) $. 
 For the magnetic $N \rightarrow \Delta $ transition form factor $G^{*}_{M}(t)$,  in the large $N_{c}$  limit,
the relevant $GPD_{N\Delta}$ can be expressed in terms of the isovector GPD
    yielding the sum rule \cite{Guidal}
\ba
  G^{*}_{M} (t) = \frac{G^{*}_{M} (t=0)}{k_{v} } \int_{-1}^{1} dx ( E_{u}(x,\xi,t) - E_{d}(x,\xi,t) )
\ea
  where $k_{v}=k_{p} - k_{n} =3.70 $
 \ba
  E_{u}(x,\xi=0,t) = d(x) e^{ [ 2 \alpha_{1} (\frac{ (1-x)^{p_{1} } }{ (x_{0} + x  )^{p_{2}} } ) ]}; \ \
  E_{d}(x,\xi=0,t) = d(x) e^{ [ 2 \alpha_{1} (\frac{ (1-x)^{p_{1} k_{d} } }{ (x_{0} + x  )^{p_{2}} } + d x (1-x) t ) ]}.
  \ea
   There are two different conventions - Jones-Scadron and Ash convention.
  They are related as 
$G^{*}_{M, J-S}  (Q^{2}) = G^{*}_{M, Ash} \sqrt{1 +\frac{Q^{2} }{ (M+m)^{2} } } $.
We will be present the data in the Ash convention.

  The results of our calculations are presented in Fig.1 (a). The  experimental data  exist up to
   $-t =8 $ GeV$^2$ and our results  show a sufficiently good coincidence with experimental data.
   It is confirmed that the form of the momentum transfer dependence of  $E(x,\xi,t)$ determined in our model  is right.

\section{The Compton cross sections}

\begin{figure}
\includegraphics[width=.3\textwidth]{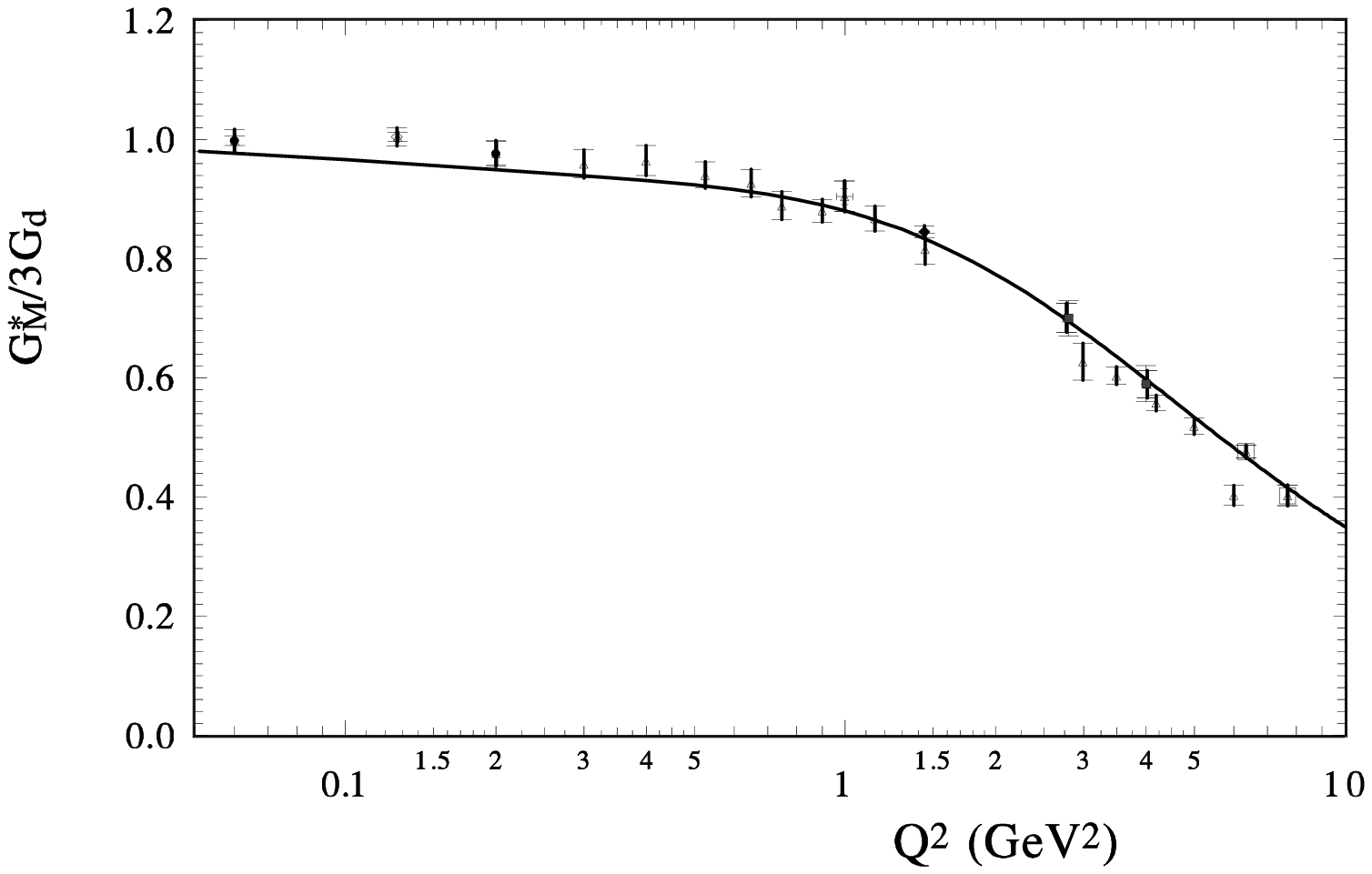} 
\includegraphics[width=.3\textwidth]{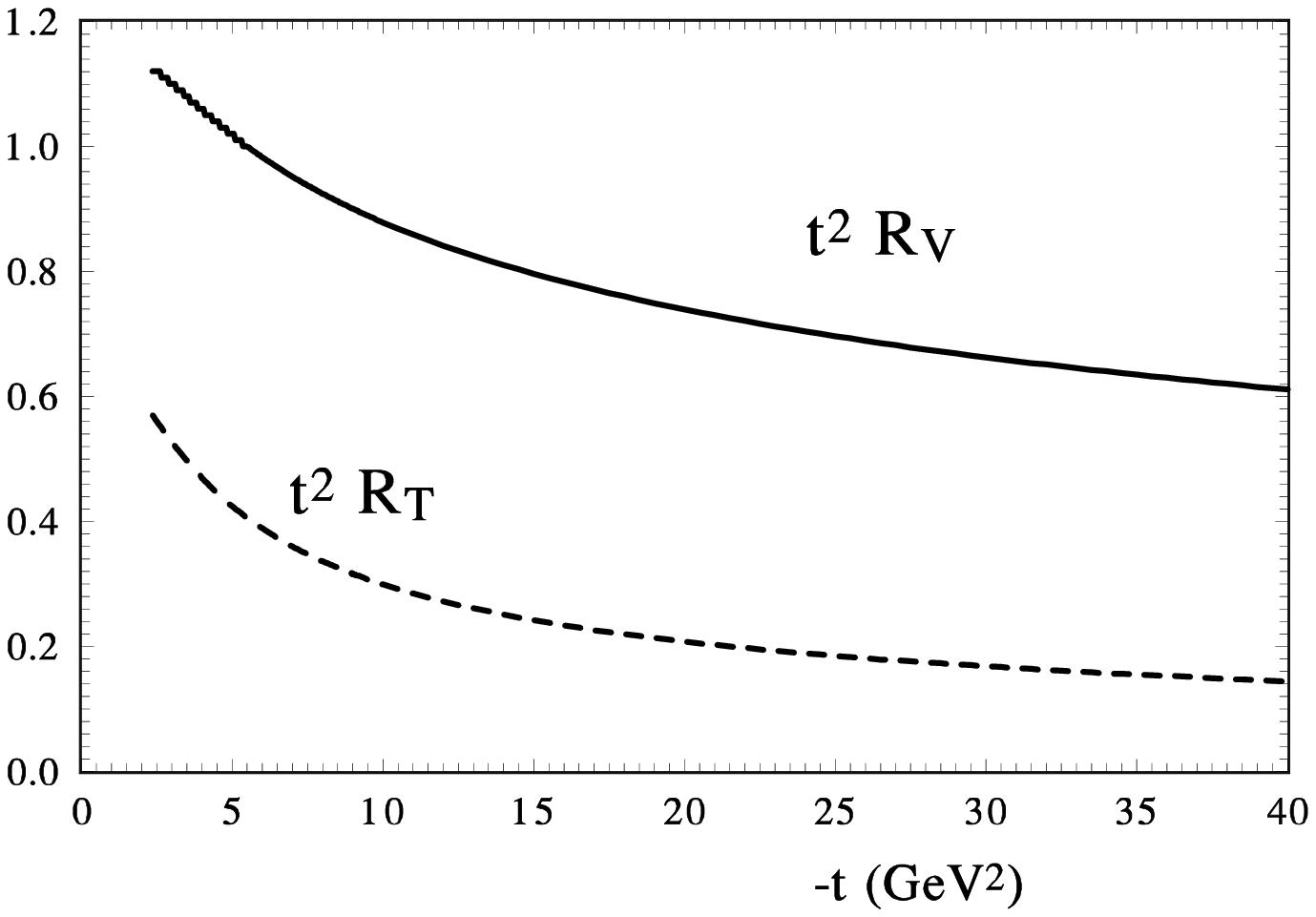} 
\includegraphics[width=.3\textwidth]{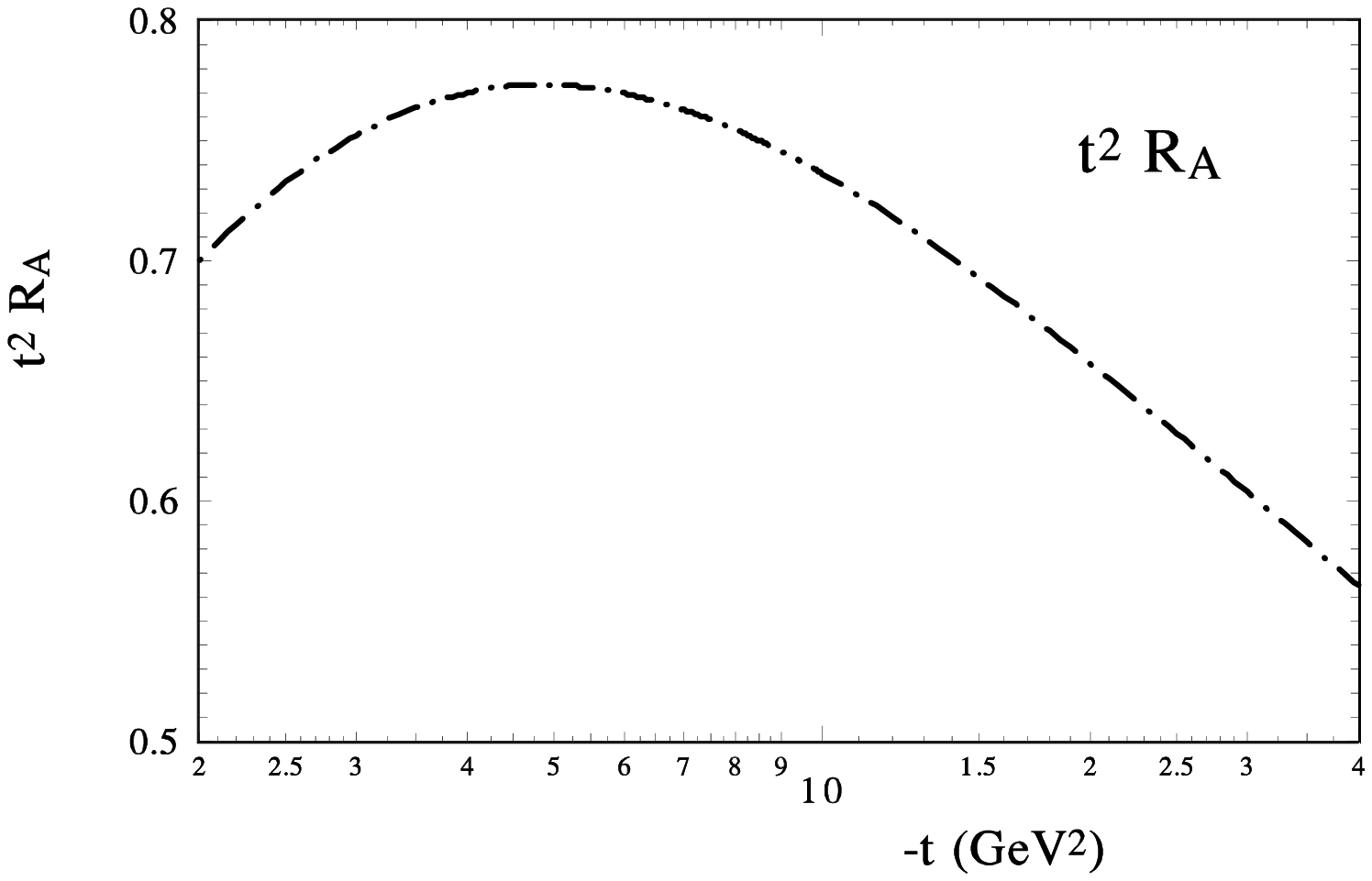} 
\caption{a) [left] The transition magnetic  form factors  $G^{*}_{M, Ash}$
(line- our calculations, points the experimental data \cite{G-data};
 b) [middle]
 the Compton form factors 
  $t^2 R_{V}(t)$ and  $t^2 R_{T}(t)$ , 
 c) [right] $t^2 R_{A}(t)$
  }
\label{Fig_1}
\end{figure}

     Now let us calculate the moments of the GPDs with inverse power of $x$. It  gives us the
     Compton form factors.
   Using the obtained form factors the reaction of the real Compton scattering
    can be calculating.
    The differential cross section for that reaction can be written as   \cite{DK-13} 
  \ba
  \frac{d\sigma}{dt} =  \frac{\pi \alpha^{2}_{em}}{s^{2}} \frac{(s-u)^{2}}{-u s}
  [R_{V}^{2}(t) \ - \ \frac{t}{4 m^{2}} R^{2}_{T}(t) 
    + \frac{t^{2}}{(s-u)^{2}} R^{2}_{A} (t)],
    \label{RCS}
\ea
  where $R_{V}((t)$, $R_{T}(t)$, $R_{A}(t)$ are the form factors given by the $1/x$
  moments of the corresponding GPDs $H^{q}(x,t)$,  $E^{q}(x,t)$, $\tilde{H}^{q}(x,t)$ .
     The last is related with the axial form factors.
     As noted in \cite{DK-13}, this factorization,
  which bears some similarity to the handbag factorization of DVCS,
  is formulated in a symmetric frame where the skewness $\xi=0$.
   For  $H^{q}(x,t)$,  $E^{q}(x,t)$ we used the PDFs obtained from the
   works \cite{Kh12} with the parameters
   obtained in our fitting procedure of the description of the proton and neutron electromagnetic
   form factors   in \cite{GPD-PRD14}.
    \ba
 R_{i}(t) =  \sum_{q} e^{2}_{q} \int_{0}^{1} \frac{dx}{x} {\cal{F}}j_{q}(x,\xi=0,t),
\ea
 where ${\cal{F}}j_{q}$ are equal $H_{q}$, $E_{q}$ and $\tilde{H}_{q}$ and give the the form factors $R_{V}(t)$, $R_{T}(t)$, $R_{A}(t)$, 
 respectively.  

  In the present work  for $\tilde{H}^{q}(x,t)$ we take $\Delta q^{e}$ in the form \cite{Khang-16}
 for  NNLO $Q_0=2$ GeV$^2 $
\ba
 x \Delta_{q}(x,Q_{0}) = N_{q} \eta_{q} x^{a_{q} } (1-x)^{b_{q} } (1 + c_{q} x);
\ea
 Assuming $SU(3)$ flavor symmetry of $\Delta \bar{q} $  the coefficient $N_{q}$ is determined as
$ \frac{1}{N_{q}} = (1+c_{q} \frac{a_{q}}{1+a_{q}+b_{q}} ) \ B(a_{q},b_{q}+1)$, 
 where $\ B(a_{q},b_{q}+1)$ is determined by
 $B(a,b)= \Gamma(a) \Gamma(b)/\Gamma(a+b) = \int_{0}^{1} t^{a-1} (1-t)^{b-1} dt $.
          The results of our calculations of the Compton form factors are shown in Fig. 1(b,c).
          Obviously, $R_{V}(t)$ and $R_{T}(t)$ have a similar momentum transfer dependence, but essentially differ in size.
  On the contrary, the axial form factor $R_{A}$ has an essentially different $t$ dependence.
          The results for the cross sections are presented in Fig.2 (a). Except the very large angles at low energy
          the coincidence with the experimental data is sufficiently good.
          The calculations of $R_{i}$  on the whole, correspond to the calculations \cite{DK-13}.

\begin{figure}
\begin{center}
\includegraphics[width=.3\textwidth]{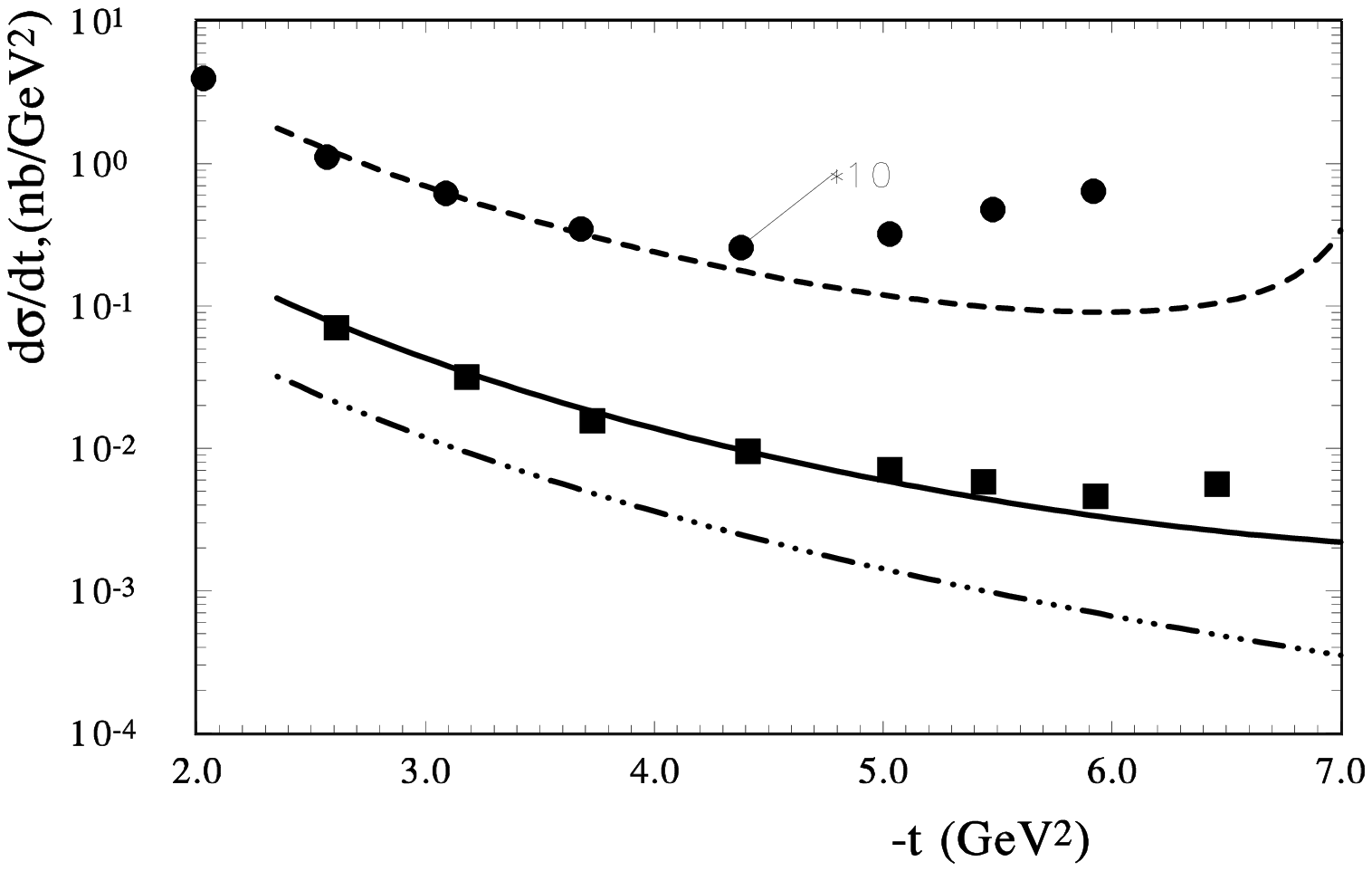} 
\includegraphics[width=.3\textwidth]{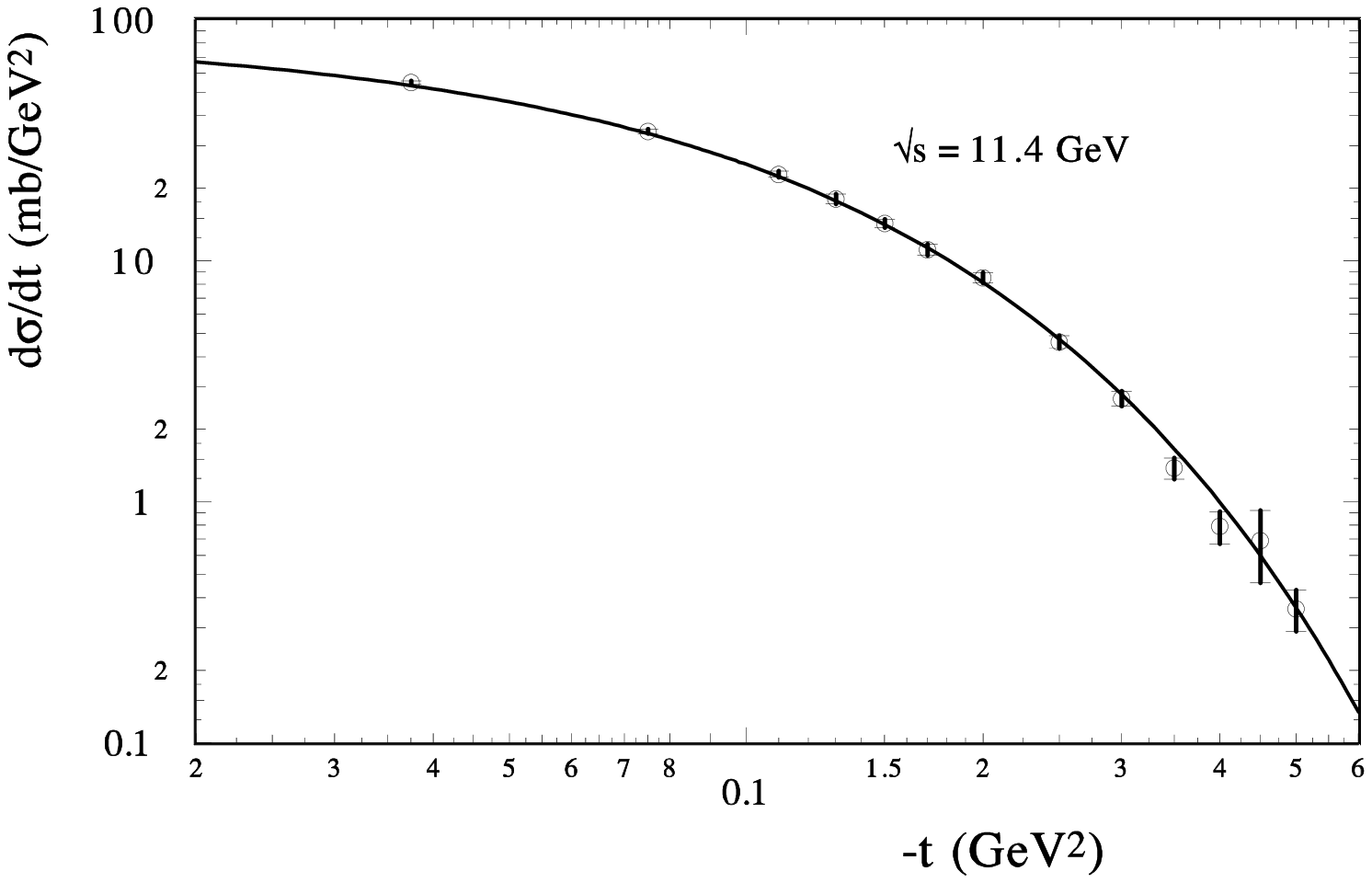} 
\includegraphics[width=.3\textwidth]{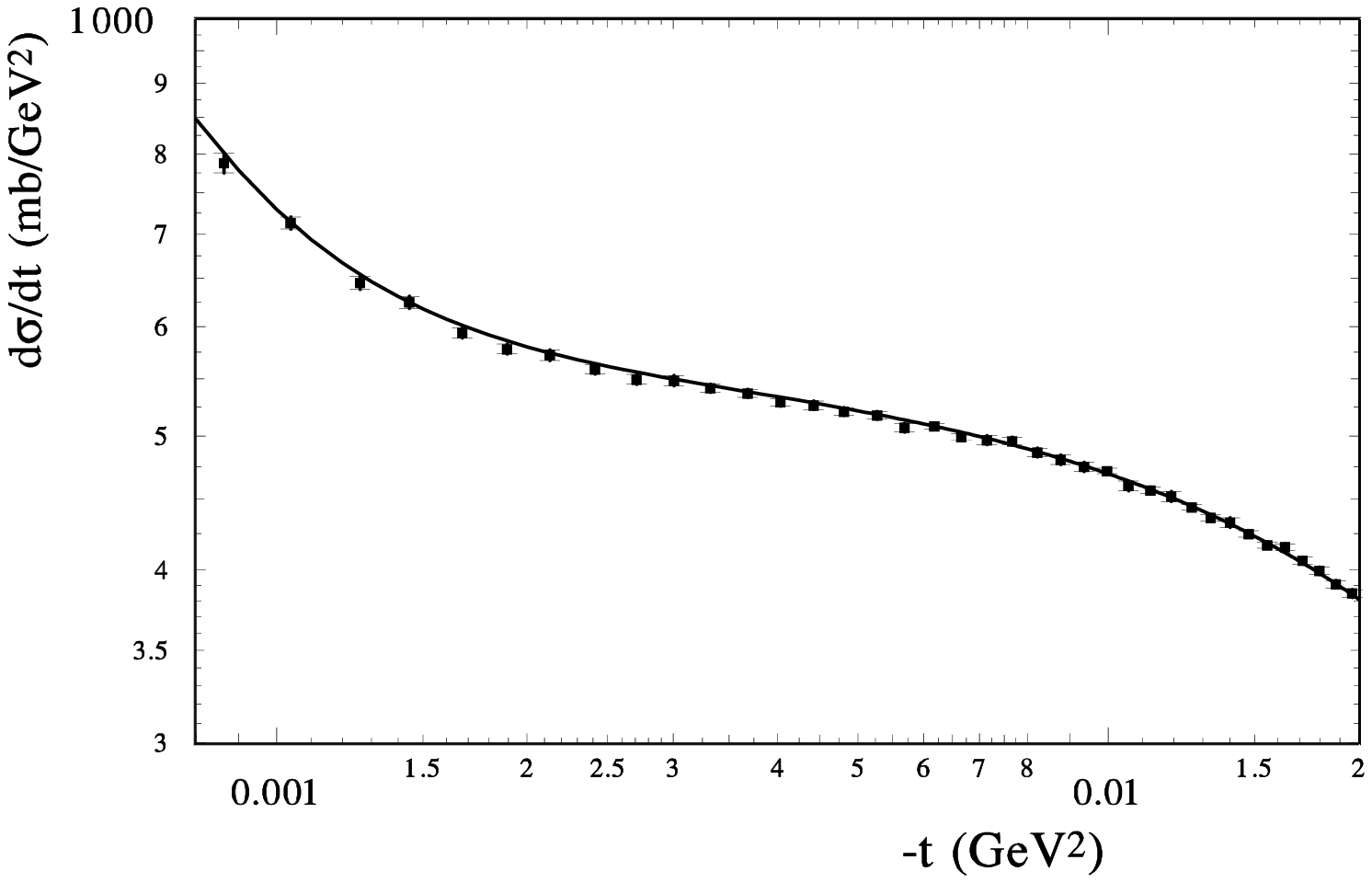} 
\caption{a) [left] Differential Compton cross sections 
 (the lines are our calculations at
$s=8.9$ GeV$^2$,
  $s=10.92$ GeV$^2$  
 and $s=20$ GeV$^2$
   the data points are for $s=8.9$ GeV$^2$ (circles);
  $s=10.92$ GeV$^2$ (squares) \cite{105-Dan07}.
 The differential cross sections of the elastic $pp$ scattering
b) [middle] at $\sqrt{s}=11.4$ GeV  
c) [right] at $\sqrt{s}= 13$ TeV;
 (lines are the calculation in the framework of the HEGS model; points - the experimental data \cite{data-Sp};
 and data of the TOTEM   Collaboration   \cite{Blois17} 
  }
 \end{center}
\label{Fig_2}
\end{figure}

\section{Hadron form factors and elastic nucleon-nucleon scattering}

        Both hadron  electromagnetic and gravitomagnetic form factors were used 
        in the framework of the high energy generalized structure
        model  of the elastic nucleon-nucleon scattering.   
        This allows us to build the model with minimum fitting
        parameters \cite{HEGS0,HEGS1,NP-hP}.

   The Born term of the elastic hadron amplitude can now be written as
  \begin{eqnarray}
 F_{h}^{Born}(s,t)=&&h_1 \ G^{2}(t) \ F_{a}(s,t) \ (1+r_1/\hat{s}^{0.5})
     +  h_{2} \  A^{2}(t) \ F_{b}(s,t) \     \\
     && \pm h_{odd} \  A^{2}(t)F_{b}(s,t)\ (1+r_2/\hat{s}^{0.5}),  \nonumber
    \label{FB}
\end{eqnarray}
  where $F_{a}(s,t)$ and $F_{b}(s,t)$  have the standard Regge form: 
$ F_{a}(s,t) \ = \hat{s}^{\epsilon_1} \ e^{B(\hat{s}) \ t}$;
$ F_{b}(s,t) \ = \hat{s}^{\epsilon_1} \ e^{B(\hat{s})/4 \ t}, $
 $   \hat{s}=s \ e^{-i \pi/2}/s_{0}$ ;  $s_{0}=4 m_{p}^{2} \ {\rm GeV^2}$, and
  $h_{odd} = i h_{3} t/(1-r_{0}^{2} t) $.
  The intercept $1+\epsilon_{1} =1.11$ was chosen from the data of the  different reactions and was fixed by the same size for all terms of the Born
  scattering amplitude.
 The slope of the scattering amplitude has the standard logarithmic dependence on the energy
 $   B(s) = \alpha^{\prime} \ ln(\hat{s}) $
  with $\alpha^{\prime}=0.24$ GeV$^{-2}$  and with some small additional term \cite{HEGS1}, 
    which reflects the small non-linear behavior of  $\alpha^{\prime}$ at small momentum transfer \cite{Sel-Df16}.
The final elastic  hadron scattering amplitude is obtained after unitarization of the  Born term.
    So, at first, we have to calculate the eikonal phase
 $ \chi(s,b) \   =  -\frac{1}{2 \pi}
   \ \int \ d^2 q \ e^{i \vec{b} \cdot \vec{q} } \  F^{\rm Born}_{h}(s,q^2)  $
  and then obtain the final hadron scattering amplitude
    \begin{eqnarray}
 F_{h}(s,t) = i s
    \ \int \ b \ J_{0}(b q)  \ \Gamma(s,b)   \ d b\, ; \ \  \ \ \ \
   {\rm with }  \ \ \ \   \Gamma(s,b)  = 1- \exp[ \chi(s,b)] .
\end{eqnarray}
     At large $t$  our model calculations are extended up to $-t=15 $ GeV$^2$.
  We added a small contribution of the energy  independent  part
  of the spin flip amplitude in the form similar to the proposed    in \cite{Kuraev-SF} and analyzed in \cite{W-Kur}.
  $ F_{sf}(s,t) \ =  h_{sf} q^3 F_{1}^{2}(t) e^{-B_{sf} q^{2}}$.
  The model is very simple from the viewpoint of the number of fitting parameters and functions.
  There are no any artificial functions or any cuts which bound the separate
  parts of the amplitude by some region of momentum transfer.
        In the framework of the model the description of the experimental data was obtained simultaniously
        at the large momentum transfer and in the Coulomb-hadron region in the energy region from $\sqrt{s}=9 $ GeV
        up to LHC energies. Figure 2(b) represents the description at  $\sqrt{s}=11.4 $ GeV and Figure 2(c)
  represents the model predictions for $\sqrt{s}=13 $ TeV, which coincide well with the preliminary data
        presented at the conference BLOIS-17 (2017) \cite{Blois-17}.

\section{Nucleon structure and radii }

Let us now consider, as an example, the neutron structure in the impact parameter  
 \cite{BurkT,RayT} representation.
We are particularly motivated by recent discussion of the definition of charge density of the  neutron
at small impact parameters corresponding to the  "center" of the neutron \cite{BurkZ,MillerZ}.
     In \cite{MillerZ}, the charge density of the neutron is related to
  $F^n_1(t)$ and calculated using
  the phenomenological representation of  $G^n_E(t)$ and $G^n_M(t)$.
 It differs essentially from the definition of the neutron charge
  distribution in the Breit frame. 
  \ba
 \rho_{G_{E}}(b)
 & = & \int d^{2}q [F_1 (q^2)+ \tau F_{2}(q^2)] \ e^{i \vec{q} \vec{b} } \nonumber \\
 & = & \sum_{q} e_{q} \int^{1}_{0}  dx \int d^{2}q [ H_{q}(x,\xi=0,q^2)
 +  \tau E_{q}(x,\xi=0,q^2)] e^{i \vec{q} \vec{b} } 
\ea
 with , as usual, $\tau = (q / m_{p})^2 $.
 Using our model of $t$-dependence of  GPDs, we may calculate
  both forms of the neutron charge distribution in the impact parameter representation
  and, moreover, determine  separate  contributions of $u$ and $d$ quarks.
 The respective
  separate contributions of $u$ and $d$-quarks are shown in Fig. 3(a).
  We can see that $u$-quarks have the large negative charge density in the centre of the
 neutron.

Let us compare the distribution of the electric charge and matter (that is, gravitational charge) in the nucleon 
with the matter density
\ba
 \rho_{0}^{Gr}(b) = \frac{1}{2 \pi}
\int^{0}_{\infty} \  dq \ q  \ J_{0}(q b) A(q^2).
\ea
 The difference of the matter density and the charge density  
 for proton is shown in Fig.3(b).
The radius of the hadron can be determined by the slope of its form factor
 $ \bigl<r^{2}_{h}\bigr>  = - 6/f_{h}(0) d/dt [f_{h} (t)]_{t=0}$.
   For the electromagnetic form factor with our form of $\tilde{G}(t)$ we have
\ba
     - \frac{6}{\tilde{G}(0)} \frac{d}{dt} [\tilde{G}(0)]     =
    {2 a_{2} t - (2 (1/L^2 + a_{1}/(2 \sqrt{t})))/(1 + a_{1} \sqrt{-t} - t/L^2)^3}
\ea
 As the coefficients $a_{i}$ and  $t$ are very small eq.(13) gives us the simple relations
 $\bigl<r^{2}_{h}\bigr>  \approx 1/L^{2}$.
   The electromagnetic form factors and the gravitomagnetic form factors  are describe almost the same form
    - the dipole form factor but with different sizes of the $L_{i}^{2}$.
    Hence we have for the electromagnetic radius
 the ratio of the radiuses of the  electromagnetic  and the gravitomagnetic form factors -
   $r_{E}/r_{A} = (1/L^{2}_{E})/(1/L^{2}_{A}) \approx 1.5$.
    It is shown that the matter distribution is concentrated in the center of the hadron and
      at large distances the charge distribution has a longer tail than the matter distribution.

\begin{figure}
\begin{center}
\includegraphics[width=.4\textwidth]{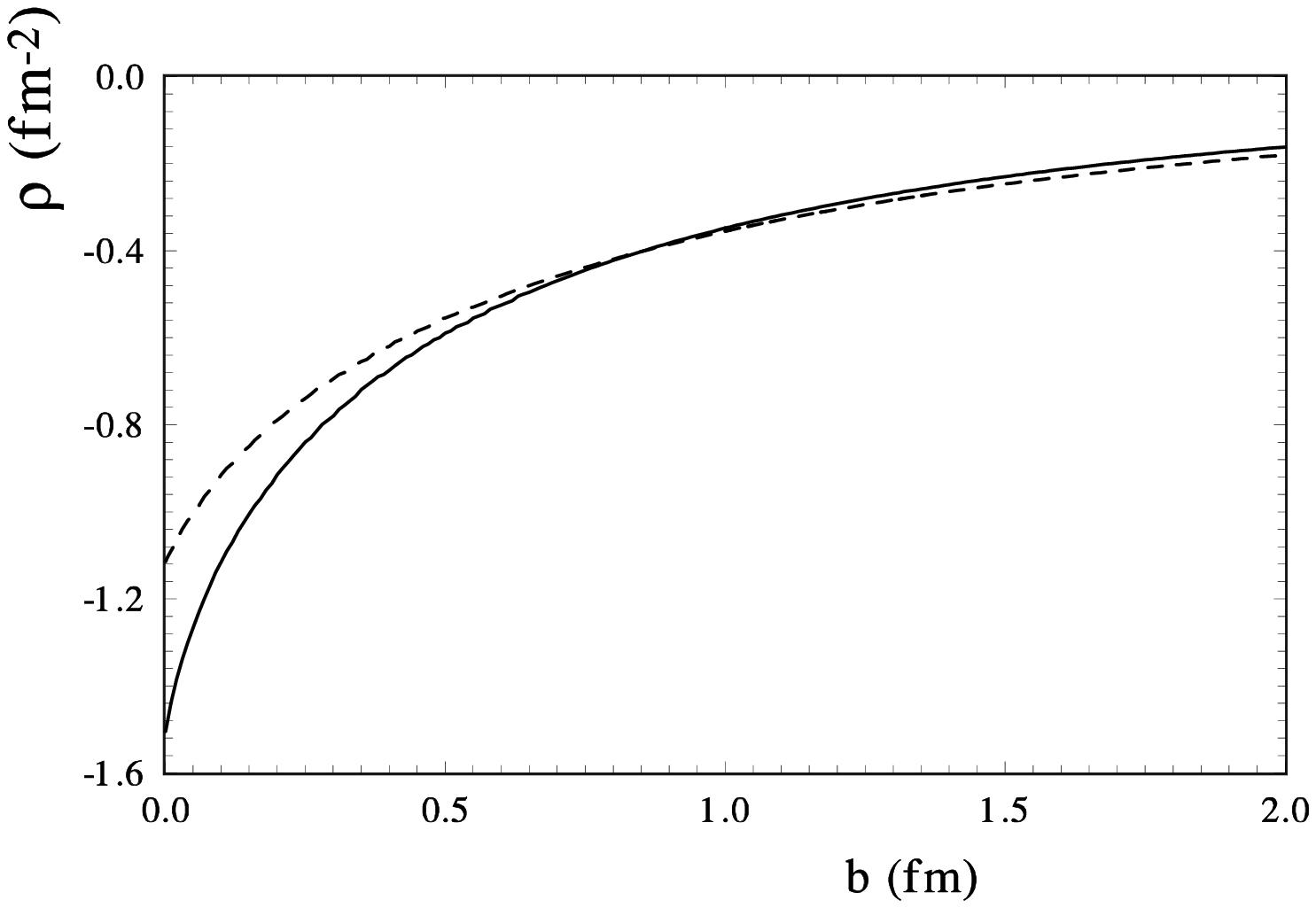} 
\includegraphics[width=.3\textwidth]{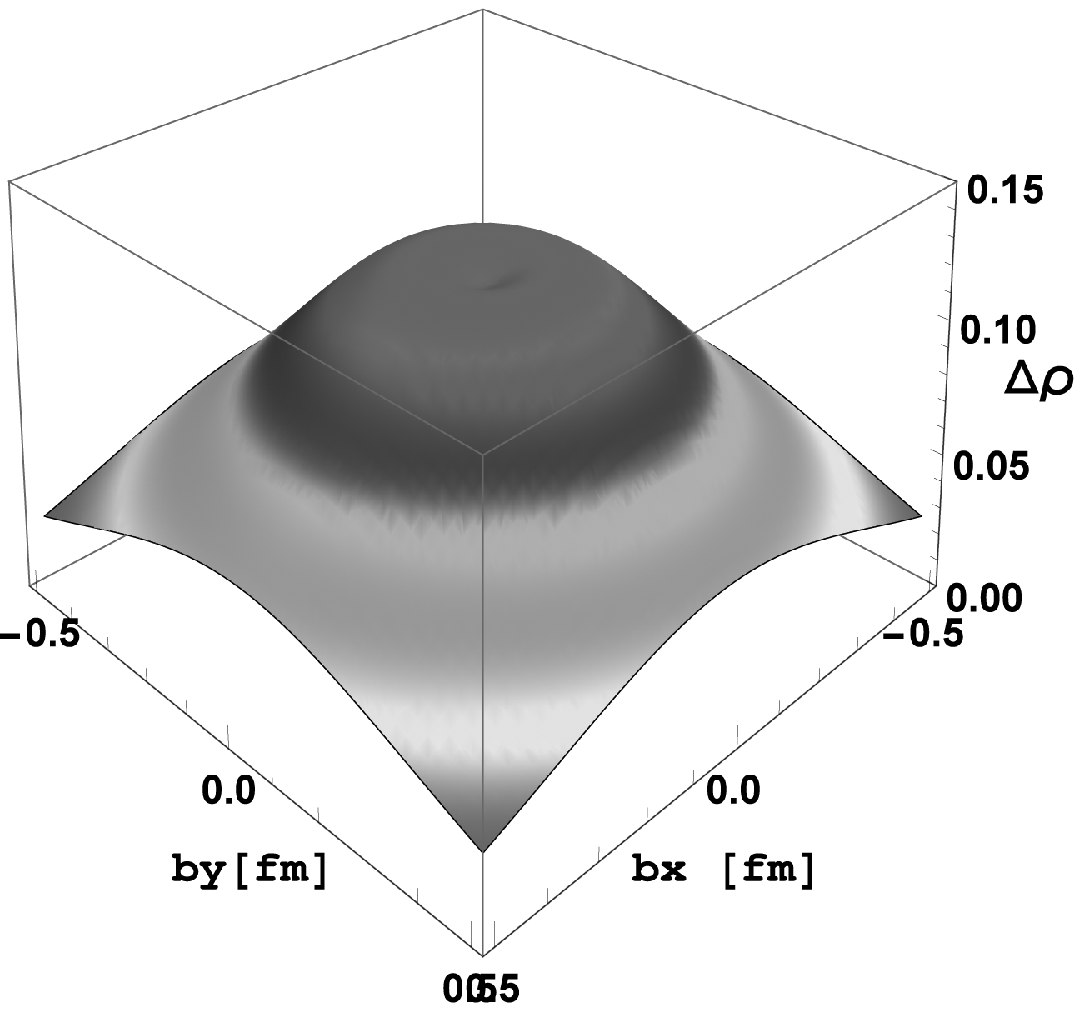} 
\caption{a) [left] $u$-quark (hard line) and $-d$-quark (dashed line) density of the neutron.
}
\end{center}
\label{Fig_}
\end{figure}

\section{Conclusion}
 The obtained new momentum transfer dependence of GPDs, 
    based on the analysis of  practically all existing experimental data on the
    electromagnetic form factors of the proton and neutron, shows a good description of 
   the $t$ dependence of a wide circle of  different reactions.
As a result, the description of various reactions is based on the same 
representation of the hadron structure.
                      Especially it concerns the  real Compton scattering and
                     high energy elastic hadron scattering.
  The new high energy generelized structure (HEGS) model, based on the electromagnetic and gravitomagnetic form factors,
   gives a good quantitative description of the existing experimental data
  of the proton-proton and proton-antiproton elastic scattering
  in a wide region of the energy scattering and momentum transfer, including the Coulomb-hadron interference region,
  with the minimum number of the fitting parameters.
    Their predictions coincides well with the new experimental data obtained at the LHC at $\sqrt{s}=7$ and $\sqrt{s}=8$ TeV
   The predictions of the model based on the obtained
      electromagnetic and  gravitomagnetic form factors
       coincides well with the recent preliminary data at $\sqrt{s}=13$ TeV \cite{Blois17}.
  The investigation of the nucleon structure  shows that the density of the matter is
   more concentrated than the charge density in the nucleon.

\vspace{0.5cm}
{\bf Acknowledgments}
 {\it The authors would like to thank J.-R. Cudell and O.V. Teryaev
   for fruitful   discussion of some questions   considered in the paper.}


\section*{References}
\begin{thebibliography}{9}
\bibitem{Rev-LHC} R. Fiore, L. Jenkovszky, R. Orava, E. Predazzi, A. Prokudin, O. Selyugin,
                    Mod.Phys., A24  (2009) 2551. 


  \bibitem{Meis-08} S. Meissner, A. Metz, M. Schlegel, and K. Goeke,
    JHEP, {\bf 0908} (2009) 056.

  \bibitem{Lorce-13} C. Lorce and B. Pasquini,
          JHEP, {\bf 1309} (2013) 138.

  \bibitem{Burk-15} M. Burkardt and B. Pasquini,
        EPJA Special Issue on "3D Structure of the Nucleon"; EPJ,


\bibitem{Mil94} D. Muller {\it et al.}, Fortsch. Phys. {\bf 42}, (1994) 101;
 \bibitem{Ji97} X.D. Ji, Phys. Lett. {\bf 78} , (1997) 610; Phys. Rev D {\bf 55} (1997) 7114;

   \bibitem{R97}   Radyushkin, A.V.,  Phys. Rev. D {\bf 56},  5524 (1997).
\bibitem{Liuti1} G.R. Goldstein, J.O. Hernandez, S. Liuti,  Phys.Rev. {\bf D84} 034007 (2011).


%
 \bibitem{Kroll04}  M.Diehl {\it et al.},  Eur.Phys. J. C  {\bf 39} (2005) 1.


\bibitem{ST-PRDGPD}  O. Selyugin, O. Teryaev, Phys. Rev.
  {\bf D 79} 033003 (2009); 



 \bibitem{GPD-PRD14}  O.V. Selyugin,
       Phys. Rev. {\bf D 89} 093007 (2014) . 



      \bibitem{DK-13}M. Diehl and P. Kroll,
          Eur.Phys.J. {\bf C73} 2397 (2013).

\bibitem{ABM12}  S. Alekhin, J. Blu"mlein, and S. Moch, Phys.Rev. D86
, 054009 (2012). 

\bibitem{Guidal} M. Guidal, M. V. Polyakov, A. V. Radyushkin, M. Vanderhaeghen,
               PhysRevD. {\bf D72} (2004) 054013.




         \bibitem{Kh12}  H. Khanpour {\it et al.}, arXiv:1205.5194


\bibitem{Khang-16}     F.Taghavi-Shahri, H. Khanpour .. 1603.03157

      \bibitem{HEGS0} O.V.~Selyugin,
      Eur.Phys.J. {\bf C72}, 2073 (2012).


 \bibitem{HEGS1}  O.V. Selyugin,
       Phys. Rev. {\bf D 91 } 113003 (2015) . 

      \bibitem{NP-hP}  O.~V.~Selyugin,
         Nucl.Phys. A {\bf 903} 54 (2013). 

 \bibitem{Sel-Df16}     O.V. Selyugin,  "Diffraction (2016)




\bibitem{G-data} F. Hagelstein, arxiv: 1710.00874.

\bibitem{105-Dan07} A. Danagoulian, et. al. (Jefferson Lab Hall A Collaboration),
 Phys.Rev.Lett., {\bf 98} 152001 (2007).




\bibitem{Kuraev-SF} M.V. Galynskii, E.A. Kuraev, Phys.Rev. D, {\bf 89} (2014) 054005.

 \bibitem{W-Kur}  O.V. Selyugin, Particel Nucleii Letters





\bibitem{BurkT} M. Burkhardt, [hep-ph]/0509316.


\bibitem{RayT} H. Dahiya, A. Mukherjee, S. Ray, [hep-ph]/0705.3580.


\bibitem{BurkZ} M. Burkardt, [hep-ph]/0709.2966v2(October, 2007).

\bibitem{MillerZ} G. A. Miller,
  Phys.Rev.Lett., {\bf 99} (2007) 112001.




\bibitem{data-Sp}
http://durpdg.dur.ac.uk/hepdata/reac.html.














    \bibitem{Blois17} M. Deile  (TOTEM Collaboration), talk in the  Workshop
          on Diffraction in High-Energy Physics, Praha (Czech. Resp.), June 26-30 (2017).


\end{thebibliography}
\end{document}